\documentclass[journal,10pt]{IEEEtran}

\ifCLASSINFOpdf
\else
   \usepackage[dvips]{graphicx}
\fi
\usepackage{url}

\usepackage{tabularx}
\usepackage{booktabs}
\usepackage{graphicx}
\usepackage{subfigure}
\usepackage{cite}
\usepackage{tikz}
\usetikzlibrary{calc,positioning}

\colorlet{dark green}{green!70!black}
\colorlet{light blue}{blue!30!white}
\colorlet{dark red}{red!50!black}

\newlength{\distance}
\newlength{\plotheight}
\newcommand{\bftab}{\fontseries{b}\selectfont}
\begin{document}

\title{Temporal Envelope and Fine Structure Cues for \\ Dysarthric Speech Detection Using CNNs}

\author{Ina Kodrasi, \IEEEmembership{Senior Member, IEEE}
\thanks{This work was supported by the Swiss National Science Foundation project no CRSII5\_173711 on ``Motor Speech Disorders: characterizing phonetic speech planning and motor speech programming/execution and their impairments''. }
\thanks{I. Kodrasi is with the Idiap Research Institute, Martigny, 1920 Switzerland (e-mail: ina.kodrasi@idiap.ch).}
}

\markboth{Vol. xx, No. xx, xxxx xx}
{Kodrasi: Bare Demo of IEEEtran.cls for IEEE Journals}
\maketitle

\begin{abstract}
    Deep learning-based techniques for automatic dysarthric speech detection have recently attracted interest in the research community.
  State-of-the-art techniques typically learn neurotypical and dysarthric discriminative representations by processing time-frequency input representations such as the magnitude spectrum of the short-time Fourier transform~(STFT).
  Although these techniques are expected to leverage perceptual dysarthric cues, representations such as the magnitude spectrum of the STFT do not necessarily convey perceptual aspects of complex sounds.
  Inspired by the temporal processing mechanisms of the human auditory system, in this paper we factor signals into the product of a slowly varying envelope and a rapidly varying fine structure.
  Separately exploiting the different perceptual cues present in the envelope (i.e., phonetic information, stress, and voicing) and fine structure (i.e., pitch, vowel quality, and breathiness), two discriminative representations are learned through a convolutional neural network and used for automatic dysarthric speech detection.
  Experimental results show that processing both the envelope and fine structure representations yields a considerably better dysarthric speech detection performance than processing only the envelope, fine structure, or magnitude spectrum of the STFT representation.     
\end{abstract}

\begin{IEEEkeywords}
  temporal envelope, temporal fine structure, dysarthria, Parkinson's disease, convolutional neural network
\end{IEEEkeywords}

\IEEEpeerreviewmaketitle

\section{Introduction}
\label{sec: intro}
Neurological disorders such as Parkinson's disease (PD) can cause dysarthria, resulting in disrupted speech production across different dimensions.
To detect and manage dysarthria, clinicians exploit perceptual assessments typically involving evaluation by ear of clinical-perceptual signs of dysarthric speech, e.g., articulation deficiencies, vowel quality changes, pitch variation, breathiness, or rhythm disruptions~\cite{Duffy_book_2019}. 
These perceptual evaluations are subject to the expertise of the clinician and can be time-consuming~\cite{Fonville_JN_2008}.
To complement the perceptual assessment of clinicians, objective dysarthric speech processing techniques have been proposed.
Such techniques can assist clinicians by automatically detecting the presence of dysarthria~\cite{Little_ITBE_2009, Bhati_SIP_2019, Janbakhshi_SPL_2020} or by automatically evaluating the patient's intelligibility and dysarthria severity~\cite{Kim_ITASLP_2015, Laaridh2017, Bhat_ICASSP_2017, Janbakhshi_ITASLP_2020}.

Typical automatic dysarthric speech detection techniques are based on handcrafting acoustic features aiming to characterize the clinical-perceptual signs of dysarthria~\cite{Tsanas_ITBE_2012, Sztaho_Interspeech_2015, Hemmerling_Interspeech_2016, Orozco-Arroyave_Interspeech_2015, Kodrasi_ICASSP_2019, Kodrasi_ITASLP_2019a, Kodrasi_Interspeech_2020, Liss_SLH_2010,Hernandez_IS_2020}.
Acoustic features such as jitter, shimmer, or fundamental frequency have been used to quantify impacted phonation~\cite{Tsanas_ITBE_2012, Sztaho_Interspeech_2015, Hemmerling_Interspeech_2016}.
Acoustic features such as Mel frequency cepstral coefficients and spectro-temporal sparsity measures have been used to quantify articulation deficiencies~\cite{Orozco-Arroyave_Interspeech_2015, Hemmerling_Interspeech_2016, Kodrasi_ICASSP_2019, Kodrasi_ITASLP_2019a, Kodrasi_Interspeech_2020}.
Further, the envelope modulation spectrum and durational measures of vocalic and intervocalic segments have been used to quantify rhythm disruptions~\cite{Liss_SLH_2010,Falk_ICASSP_2011, Hernandez_IS_2020}.
Although successful results have been reported using handcrafted features, such features may fail to characterize more abstract but similarly important perceptual dysarthric cues.
Consequently, there has been a growing interest in the research community to develop deep learning-based automatic dysarthric speech detection techniques~\cite{Millet_ICASSP_2019,Mallela_IS_2020, Vasquez2017,Vaiciukyna_SOTSG_2017,Kwanghoon_IS_2018, Vasquez_SC_2020, Janbakhshi_ICASSP_2021}.

In~\cite{Millet_ICASSP_2019}, raw neurotypical and dysarthric speech segments have been used to train a long short-term memory Siamese networks that learns discriminative representations.
Raw speech segments have also been used in~\cite{Mallela_IS_2020}, where convolutional neural networks~(CNNs) have been trained instead.
Given the limited amount of pathological training data, contributions exploiting raw speech segments are seldom.
Instead, mainstream techniques rely on processing the magnitude spectrum of time-frequency representations such as the Mel spectogram~\cite{Vaiciukyna_SOTSG_2017, Kwanghoon_IS_2018, Vasquez_SC_2020}, the continuous wavelet transform~\cite{Vasquez2017}, or the short-time Fourier transform~(STFT)~\cite{Vasquez2017, Vaiciukyna_SOTSG_2017, Janbakhshi_ICASSP_2021}.
Although these techniques are expected to leverage perceptual dysarthric cues, such representations do not necessarily convey perceptual aspects of complex sounds~\cite{Rosen_PT_1992}.

Within the cochlea, speech signals are filtered into a series of narrowband signals with a slowly varying envelope imposed on a rapidly oscillating carrier, i.e., the temporal fine structure.
The relative importance of the temporal envelope and fine structure to speech perception has been the subject of a wide range of literature for decades, with particular focus on the importance of these cues for speech intelligibility in the presence of interference and the effects of hearing loss on the processing of these cues in the auditory nerve~\cite{Smith_nature_2002, Liu_JASA_2006, Moore_JASA_2009, Henry_Frontiers_2014}.
Furthermore, processing the temporal envelope and/or fine structure has been crucial for applications such as automatic speech recognition or speech enhancement~\cite{Moritz_ITASLP_2015, Purushothaman_IS_2020, Thoidis_Electronics_2020}.
The importance of fine structure cues for dysarthric speech assessment has been recently demonstrated in~\cite{Gurugubelli_SC_2020}, where these cues have been extracted using a single frequency filtering representation and exploited in an i-vector based dysarthria detection system.
Although the relative importance of the temporal envelope and fine structure for speech perception is still debated~(cf., ~\cite{Shamma_JASA_2013}), it is established that envelope signals contain phonetic information as well as stress and voicing information, whereas fine structure signals are important for pitch perception and vowel quality~\cite{Rosen_PT_1992, Smith_nature_2002}.

Inspired by these temporal processing mechanisms of the human auditory system, in this paper we propose a deep learning-based dysarthric speech detection technique which separately processes the temporal envelope and fine structure signals.
Two discriminative representations separately learned from the temporal envelope and fine structure using CNNs are then exploited for automatic dysarthric speech detection.
To the best of our knowledge, the extraction of temporal fine structure signals through an auditory-inspired filter bank and their use in deep learning-based approaches has never been investigated.

Experimental results in Section~\ref{sec: res} show that the temporal envelope contains more cues for dysarthric speech detection than the temporal fine structure. 
Further, it is shown that the proposed approach which exploits cues from both signals to learn two discriminative representations provides a considerable performance increase as opposed to learning a single discriminative representation from inputs where dysarthric cues are partially lost or intermingled (such as in the magnitude spectrum of the STFT representation).

\section{Temporal Envelope and Fine Structure Dysarthric Speech Detection}
In the following, the proposed temporal envelope and fine structure (TEFS)-based dysarthric speech detection system is described.
Section~\ref{sec: model} presents the computation of the temporal envelope and fine structure representations, whereas Section~\ref{sec: cnn} presents the used CNN.

\subsection{Temporal envelope and fine structure representations}
\label{sec: model}
We denote the speech signal of a neurotypical or dysarthric speaker by $s(n)$, with $n$ being the time index.
When a clinician listens to this signal to conduct their perceptual assessment, the cochlea processes the signal through frequency analysis and temporal envelope and fine structure decomposition.
To mimic cochlear frequency analysis, we use a bank of $K$ band-pass filters to split the signal $s(n)$ into $K$ complementary frequency bands of equal width along the human basiliar membrane~\cite{Smith_nature_2002}.
Let $s_c(k,n)$ denote the subband signal at the output of the \mbox{$k$-th} band-pass filter, with $k = 1, \ldots, K$.
The subband temporal envelope and fine structure signals are computed through the analytic representation of $s_c(k,n)$, i.e.,
\begin{equation}
  \label{eq: an}
  s_a(k,n) = s_c(k,n) + j{\cal{H}}\{s_c(k,n) \},
\end{equation}
where ${\cal{H}}\{s_c(k,n) \}$ denotes the Hilbert transform of $s_c(k,n)$.
Based on~(\ref{eq: an}), the subband temporal envelope and fine structure signals $e_c(k,n)$ and $f_c(k,n)$ can be computed as
\begin{eqnarray}
  \label{eq: env}
  e_c(k,n) & =  \sqrt{ s_c^2(k,n) + {\cal{H}}^2\{s_c(k,n) \} }, \\
  \label{eq: fs}
  f_c(k,n) & = \cos \left[ \arctan \left( \frac { {\cal{H}}\{s_c(k,n)\} }{ s_c(k,n)} \right) \right].
\end{eqnarray}
These signals are then averaged within time frames of length $L_w$ to create the temporal envelope and fine structure representations $E_c(k,l)$ and $F_c(k,l)$, $l=1, \ldots, L$, with $L$ being the total number of time frames in $s(n)$.
To further emphasize these representations, log scaling is applied to obtain the final envelope and fine structure representations $E(k,l)$ and $F(k,l)$.
Since $E_c(k,l)>0$~(cf.~(\ref{eq: env})), the final envelope representation is obtained as $E(k,l) = \log_{10}E_c(k,l)$.
Since $-1 \leq F_c(k,l) \leq 1$~(cf.~(\ref{eq: fs})), the final fine structure representation is obtained as $F(k,l) =  {\rm sgn} \{F_c(k,l) \}  \log_{10}|F_c(k,l)|$ such that zero-crossings are preserved.

\begin{figure}[t]
  \centering
  \includegraphics[scale=0.38]{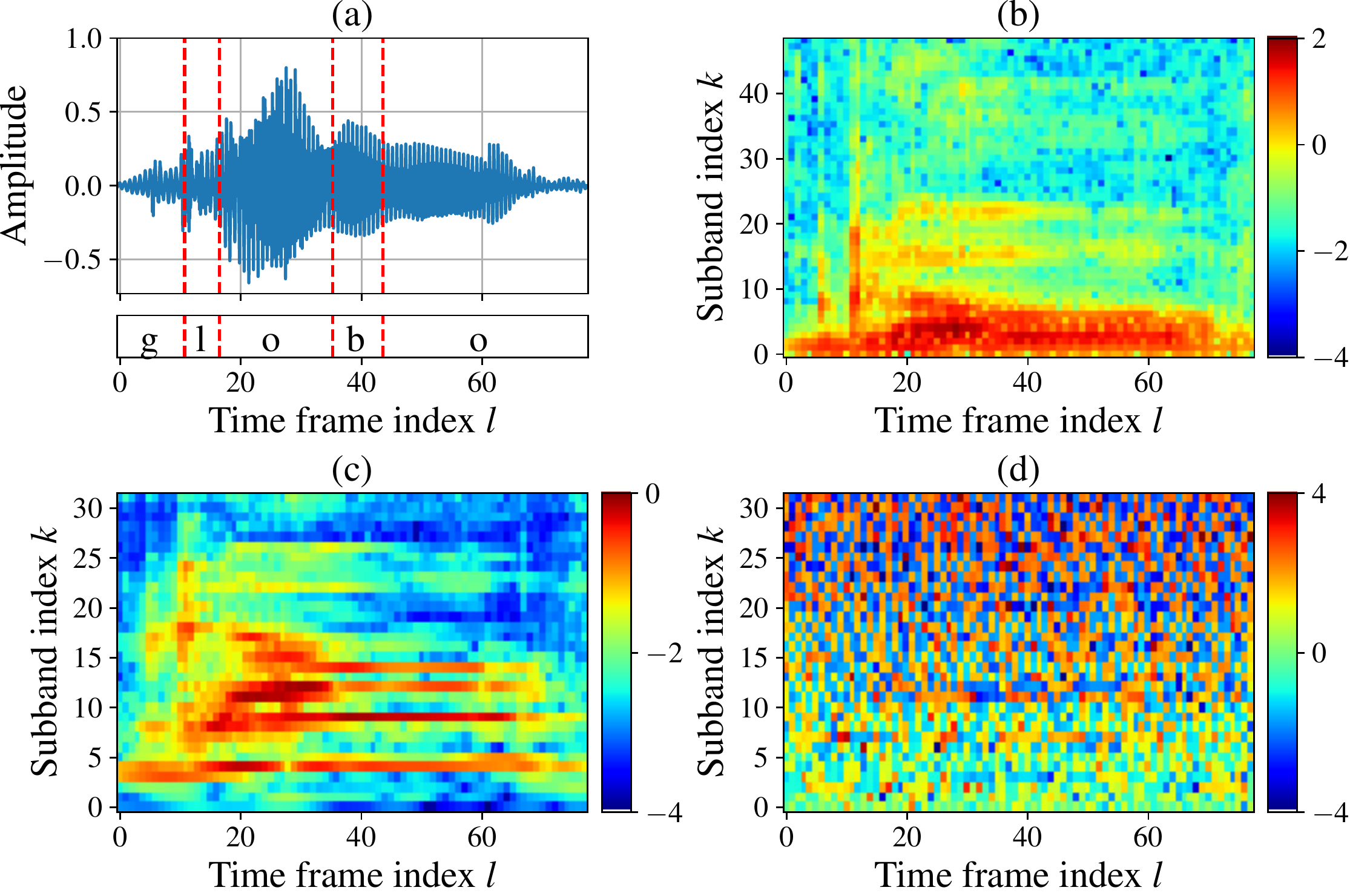}
  \caption{Different representations of the exemplary utterance \emph{globo}: (a) time domain signal $s(n)$, (b) magnitude spectrum of the STFT using $L_w = 6$~ms, (c) envelope $E(k,l)$ using $K=32$ and $L_w = 6$ms, and (d) fine structure $F(k,l)$ using $K=32$ and $L_w = 6$ms.}
\label{fig: reps}
\end{figure}

Fig.~\ref{fig: reps}(a) depicts an exemplary utterance $s(n)$ from the database described in Section~\ref{sec: data}.
The temporal envelope and fine structure representations $E(k,l)$ and $F(k,l)$ for this utterance computed using $K=32$ and $L_w = 6$~ms are depicted in Figs.~\ref{fig: reps}(c) and~\ref{fig: reps}(d).
These representations convey different perceptual cues, with the envelope representation conveying phonetic information as well as stress and voicing information and the fine structure representation conveying pitch and vowel quality information.
For completeness, the commonly used (logarithm of the) magnitude spectrum of the STFT representation of $s(n)$ using $L_w = 6$~ms is depicted in Fig.~\ref{fig: reps}(b), where these different perceptual cues are either partially lost or intermingled.\footnote{It should be noted that the STFT representation results in a trade-off between spectral and temporal resolution.
Hence, although the same window length $L_w=6$~ms is used to compute the STFT, the number of STFT subbands differs from the number of subbands used in the envelope and fine structure representations.}
This perceptual information loss occurs not only because the phase of the STFT is disregarded, but also because the STFT uses uniform filter banks which do not approximate well auditory frequency analysis.

\subsection{Convolutional neural network}
\label{sec: cnn}
\begin{figure}[t]
  \centering
    \begin{tikzpicture}[node distance=\distance,font=\scriptsize]
    \node[draw,inner sep=0pt] (spec) {\includegraphics[height=\plotheight]{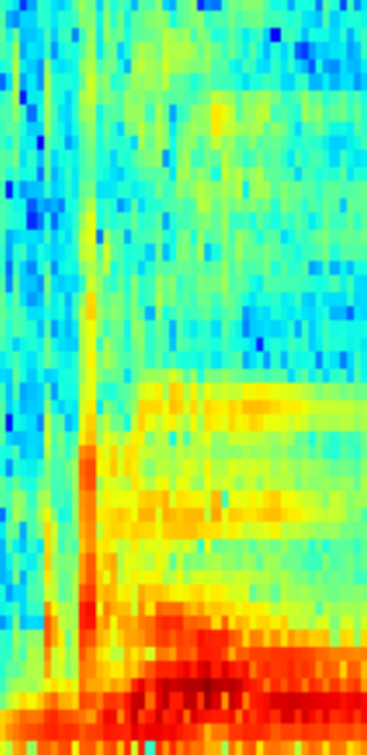}};
    \node[right=of spec,draw,align=center,fill=dark green] (2d-layers) {Conv2D layers \\ MaxPool2D layers};
    \node[right=of 2d-layers,draw,align=center,fill=light blue] (full-layers) {Fully-connected \\ layers};
    \node[right=of full-layers] (output) {0/1};

    \node[above,align=center] at (spec.north) {Input representation \\ $(K \times B)$};

    \begin{scope}[>=stealth,thick,dark red]
      \draw[->] (spec) -- (2d-layers);
      \draw[->] (2d-layers) -- (full-layers);
      \draw[->] (full-layers) -- (output);
    \end{scope}
  \end{tikzpicture}
  \caption{Block diagram of the baseline CNN-based dysarthric speech detection system from~\cite{Vasquez2017}.}
  \label{fig: baseline}
\end{figure}
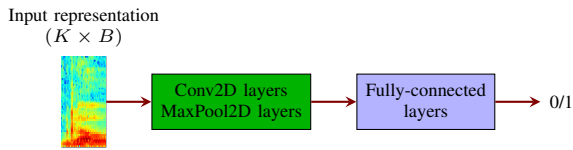

Once a signal representation is computed, the CNN depicted in the block diagram in Fig.~\ref{fig: baseline} can be trained for automatic dysarthric speech detection as in~\cite{Vasquez2017, Janbakhshi_ICASSP_2021}. 
The CNN receives as input $(K \times B)$--dimensional neurotypical and dysarthric speech representations (envelope, fine structure, STFT, or any other time-frequency representation), with $B$ being a user-defined number of time frames.
Through alternating between convolutional and pooling layers, the CNN is expected to extract robust discriminative representations of neurotypical and dysarthric speech.
These extracted representations are then exploited in fully-connected layers~(FCLs) trained to decide whether the $(K \times B)$--dimensional input representation corresponds to a neurotypical or dysarthric speaker. 
While this approach can be used on the individual envelope and fine structure representations described in Section~\ref{sec: model}, it is sub-optimal since only the cues available in one representation would be exploited~(cf. Section~\ref{sec: res}).

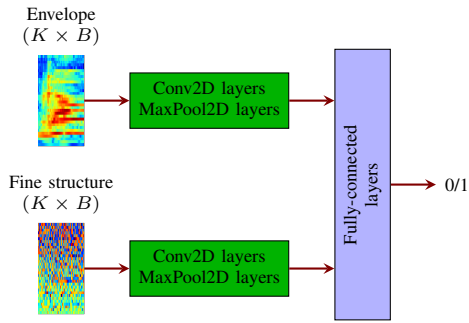
\begin{figure}[t]
  \centering
    \begin{tikzpicture}[node distance=\distance,font=\scriptsize]
    \node[draw,inner sep=0pt] (spec-1) {\includegraphics[height=\plotheight]{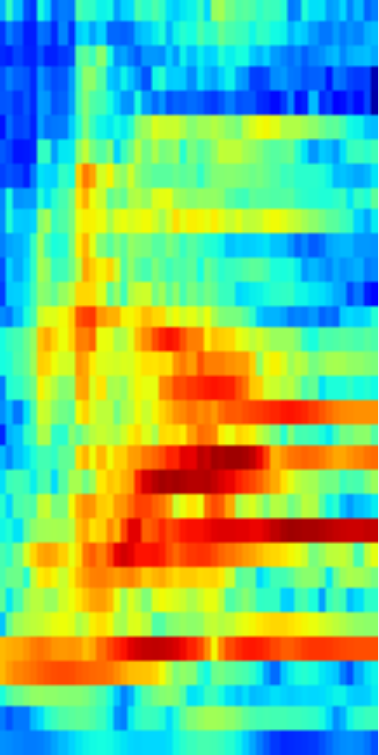}};
    \node[draw,inner sep=0pt,below=1.7\distance of spec-1] (spec-2) {\includegraphics[height=\plotheight]{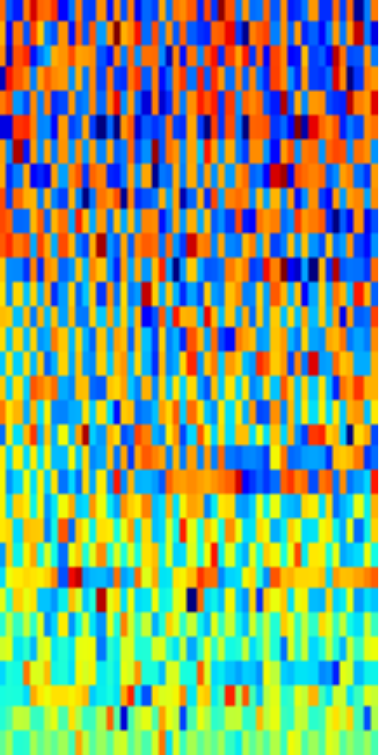}};

    \node[right=of spec-1,draw,align=center,fill=dark green] (2d-layers-1) {Conv2D layers \\ MaxPool2D layers};
    \node[right=of spec-2,draw,align=center,fill=dark green] (2d-layers-2) {Conv2D layers \\ MaxPool2D layers};

    \coordinate (m) at ([xshift=\distance] $(2d-layers-1.east)!1/2!(2d-layers-2.east)$);
    \node[below,draw,align=center,fill=light blue,rotate=90,minimum width=6\distance] at (m) (full-layers) {Fully-connected \\ layers};
    \node[right] at ([xshift=\distance] full-layers.south) (output) {0/1};

    \node[above,align=center] at (spec-1.north) {Envelope \\ $(K \times B)$};
    \node[above,align=center] at (spec-2.north) {Fine structure \\ $(K \times B)$};

    \begin{scope}[>=stealth,thick,dark red]
      \draw[->] (spec-1) -- (2d-layers-1);
      \draw[->] (spec-2) -- (2d-layers-2);
      \draw[->] (2d-layers-1) -- (2d-layers-1 -| full-layers.north);
      \draw[->] (2d-layers-2) -- (2d-layers-2 -| full-layers.north);
      \draw[->] (full-layers) -- (output);
    \end{scope}
  \end{tikzpicture}
  \caption{Block diagram of the proposed temporal envelope and fine structure-based dysarthric speech detection system.}
  \label{fig: tfs}
\end{figure}

To exploit cues available in both the temporal envelope and fine structure representations, we propose to use the TEFS-based dysarthric speech detection system depicted in Fig.~\ref{fig: tfs}.
As shown in this figure, we use individual convolutional and pooling layers that operate on the envelope and fine structure representations.
Two discriminative representations are extracted and jointly exploited in FCLs trained to detect dysarthric speech.
As shown in Section~\ref{sec: res}, such an approach yields a considerably better performance than using the system depicted in Fig.~\ref{fig: baseline} on the individual envelope, fine structure, or magnitude of the STFT representations.

\section{Material and Method}
\label{sec: exp}

\subsection{Database}
\label{sec: data}
We consider Spanish recordings of $50$ PD patients and $50$ neurotypical speakers from the PC-GITA database~\cite{GITA}.
The database is well balanced in terms of age and gender and the recordings are captured in a sound proof booth at a sampling frequency of $44.1$~kHz.
For the results presented in the following, we use recordings of $24$ different words and of a phonetically balanced text downsampled to $16$~kHz.
The average length of the available speech material for each speaker is $32.1$~s.

\subsection{Proposed TEFS-based network}
\label{sec: tefs}
The proposed TEFS-based dysarthric speech detection system depicted in Fig.~\ref{fig: tfs} operates on segments of envelope and fine structure representations.
For the results presented in this paper, these segments are computed as follows.

We design band-pass filters spanning the range from $80$~Hz to $7200$~Hz, with cut-off frequencies spaced in equal steps along the cochlear frequency map~\cite{Greenwood_JASA_1990, Smith_nature_2002}.
The number of filters used is $K = 32$.
After band-pass filtering the input signal, the envelope and fine structure representations are computed as described in Section~\ref{sec: model} using $L_w = 6$~ms.
Finally, ($K \times B$)-dimensional segments using $B = 50$ and a $50\%$ overlap  are extracted and used as inputs to the system.
Table~\ref{tbl: arch_net} summarizes the architecture of the proposed system, which has approximately $1$ million trainable parameters.
As shown in this table, the same architecture (adapted from~\cite{Vasquez2017}) is used for both the envelope and fine structure branches.

\begin{table}[t]
  \begin{center}
    \caption{Architecture of the proposed TEFS-based dysarthric speech detection technique. BN refers to batch normalization.}
    \label{tbl: arch_net}
    \def\tabcolsep{2.5pt}
    \begin{tabularx}{\linewidth}{Xl}
      \toprule
      Layers & Envelope or fine structure branch \\
      \toprule
      Input & ($K \times B$)-dimensional envelope \\
      Conv2D+ReLU+BN & in=$1$, out=$64$, kernel=$(2,2)$, stride=$(1,1)$ \\
      MaxPool2D & in=$64$, out=$64$, kernel=$(2,2)$, stride=$(2,2)$ \\
      Conv2D+ReLU+BN & in=$64$, out=$64$, kernel=$(3,3)$, stride=$(1,1)$ \\
      MaxPool2D & in=$64$, out=$64$, kernel=$(2,2)$, stride=$(2,2)$ \\
      Dropout & probability = $0.5$ \\
      \hline
      FCL+ReLU & in=$8448$, out=$128$ \\
      FCL+Softmax & in=$128$, out=$2$ \\
      \bottomrule 
   \end{tabularx}
 \end{center}
\end{table}

\subsection{Baseline networks}
To analyze the individual cues available for dysarthric speech detection in the envelope and fine structure representations, the baseline CNN depicted in Fig.~\ref{fig: baseline} is separately trained on the envelope and fine structure representations computed as described in Section~\ref{sec: tefs}.
To further demonstrate the advantages of the proposed approach, we have also trained such a baseline CNN on the magnitude of the STFT representation.

The STFT is computed using a weighted overlap-add framework with a Hanning analysis window without overlap.
As previously mentioned, the STFT yields a trade-off between spectral and temporal resolution.
For a fair comparison, we consider an STFT analysis window length $L_w = 3.875$~ms, such that the same spectral dimension (i.e., $K = 32$) is obtained as for the envelope and fine structure representations.
After computing the STFT, ($K \times B$)-dimensional segments using $B = 50$ and a $50\%$ overlap are extracted and used as inputs to the system.

Two architectures A$_1$ and A$_2$ are considered for these baseline networks.
For A$_1$, the same architecture as the one presented in Table~\ref{tbl: arch_net} for the individual envelope or fine structure branches is used.
After the dropout layer, a FCL (with an input dimension of $4224$ and output dimension of $2$) followed by the softmax function is used.
Such an architecture has approximately $45$ thousand trainable parameters.
Since A$_1$ has a considerably lower number of parameters than the proposed system in Table~\ref{tbl: arch_net}, we also consider the deeper architecture A$_2$ shown in Table~\ref{tbl: arch_b}.
This architecture has approximately $1$ million trainable parameters, comparable to the proposed system in Table~\ref{tbl: arch_net} (cf.~Section~\ref{sec: tefs}).

\begin{table}[t]
  \begin{center}
    \caption{Architecture A$_2$ for the baseline systems. BN refers to batch normalization.}
    \label{tbl: arch_b}
    \def\tabcolsep{2.5pt}
    \begin{tabularx}{\linewidth}{Xl}
      \toprule
      Layers & \\
      \toprule
      Input & ($K \times B$)-dimensional envelope \\
      Conv2D+ReLU+BN & in=$1$, out=$64$, kernel=$(2,2)$, stride=$(1,1)$ \\
      MaxPool2D & in=$64$, out=$64$, kernel=$(2,2)$, stride=$(2,2)$ \\
      Conv2D+ReLU+BN & in=$64$, out=$64$, kernel=$(3,3)$, stride=$(1,1)$ \\
      MaxPool2D & in=$64$, out=$64$, kernel=$(2,2)$, stride=$(2,2)$ \\
      Conv2D+ReLU+BN & in=$64$, out=$64$, kernel=$(4,4)$, stride=$(1,1)$ \\
      MaxPool2D & in=$64$, out=$64$, kernel=$(2,2)$, stride=$(2,2)$ \\
      Dropout & probability = $0.5$ \\
      \hline
      FCL+ReLU & in=$256$, out=$4096$ \\
      FCL+Softmax & in=$4096$, out=$2$ \\
      \bottomrule 
   \end{tabularx}
 \end{center}
\end{table}

\subsection{Training and evaluation}
The evaluation strategy is a speaker-independent stratified $10$-fold cross-validation, ensuring that each fold is balanced in terms of gender and in terms of the number of neurotypical and PD speakers.
In each training fold, a development set with the same size as the test set is used for early-stopping.
Z-score normalization is applied to all input representations and networks are trained using the stochastic gradient descent algorithm and the cross-entropy loss.
The batch size is $128$ and the initial learning rate is $0.01$.
The learning rate is halved if the loss on the development set has not decreased for 5 consecutive iterations.
Training is stopped when the learning rate has decreased beyond $10^{-6}$ or after $100$ epochs.
The trained models output a prediction score for each of the $(K \times B)$--dimensional segments and the final decision for an unseen speaker is made by applying soft voting on these segment-level prediction scores.

The baseline CNNs trained on the envelope, fine structure, or STFT representations are randomly initialized.
The convolutional layers of the proposed TEFS-based technique are initialized with the convolutional layers of trained baseline systems, with the upper branch network in Fig.~\ref{fig: tfs} initialized with the baseline architecture A$_1$ trained on the envelope representation and the lower branch network in Fig.~\ref{fig: tfs} initialized with the baseline architecture A$_1$ trained on the fine structure representation.

Dysarthric speech detection performance is evaluated in terms of the area under ROC curve (AUC) and classification accuracy for a decision threshold of $0.5$.
To reduce the impact of initialization on the final model parameters, we have trained all networks with $5$ different random seeds.
To reduce the impact of the speaker split into training and testing folds, we have repeated this training procedure for $5$ different splits of speakers.
Hence, we have trained $250$ models for each considered network, i.e., $5$ models for each of the $10$ folds obtained using $5$ different fold splits.
The reported performance measures are the mean and standard deviation of the performance obtained across these different models.

\begin{table}[t]
  \begin{center}
    \caption{Performance using the baseline systems trained on the STFT, envelope, and fine structure representations and using the proposed TEFS-based technique.}
    \label{tbl: perf}
    \def\tabcolsep{5.5pt}
    \begin{tabularx}{\linewidth}{Xrr}
      \toprule
      Network & AUC & Accuracy [\%] \\
      \toprule
      A$_1$ - Magnitude of STFT & $0.76 \pm 0.14$ & $69.52 \pm 14.04$ \\
      A$_2$ - Magnitude of STFT & $0.79 \pm 0.14$ & $69.76 \pm 13.71$ \\
      \hline
      A$_1$ - Envelope & $0.83 \pm 0.14$ & $73.80 \pm 11.75$ \\
      A$_2$ - Envelope & $0.81 \pm 0.13$ & $70.50 \pm 11.42$ \\
      \hline
      A$_1$ - Fine structure & $0.72 \pm 0.15$ & $65.68 \pm 12.38$ \\
      A$_2$ - Fine structure & $0.66 \pm 0.15$ & $61.40 \pm 13.36$ \\
      \hline
      TEFS & \bftab 0.93 $\pm$ 0.08 & \bftab 85.72 $\pm$ 10.38  \\
      \bottomrule
   \end{tabularx}
 \end{center}
\end{table}

\section{Results}
\label{sec: res}
Table~\ref{tbl: perf} presents the performance obtained using the baseline CNNs trained on different input representations and using the proposed TEFS-based technique.

It can be observed that using A$_2$ for any of the baseline systems typically yields a lower performance than using A$_1$.
Such a result can be explained by the considerably larger number of parameters in A$_2$ in comparison to A$_1$, resulting in overfitting and poor generalization performance for A$_2$.
Further, it can be observed that out of the considered baseline systems, using the envelope representation outperforms using the magnitude spectrum of the STFT or the fine structure representation.
These results show  that the envelope of a signal contains more cues for dysarthric speech detection than its fine structure.
Further, these results confirm the advantages of using these auditory-inspired representations for CNN-based dysarthric speech detection.



Finally, Table~\ref{tbl: perf} shows that the proposed TEFS-based technique yields a better performance than all considered baseline systems for both performance measures, with an AUC of $0.93$ an accuracy score of $85.72\%$.
These results show that although dysarthric cues can be more prominent in the envelope than in the fine structure, exploiting both representations is very beneficial for deep learning-based dysarthric speech detection.

\section{Conclusion}
In this paper we have proposed a deep learning-based dysarthric speech detection technique inspired by the temporal processing mechanisms of the human auditory system.
The proposed technique relies on decomposing speech signals into their envelope and fine structure counterparts, each containing different perceptual cues for dysarthric speech detection.
By separately processing the envelope and fine structure through individual convolutional and pooling layers, two discriminative representations are learned and jointly exploited for dysarthric speech detection.
Experimental results on a Spanish database of neurotypical and PD speakers have shown that the envelope representation contains more discriminative cues than the fine structure representation.
Further, experimental results have shown that exploiting both envelope and fine structure representations yields a considerably better dysarthric speech detection performance than exploiting only the envelope, fine structure, or STFT representation.     
In the future, we plan to investigate how the incorporation of more complex auditory models affects the extracted discriminative representations and the final performance of the system.
Further, we plan to investigate different architectures for processing the temporal envelope and fine structure cues.

\bibliographystyle{IEEEtran}
\bibliography{refs}

\end{document}